\documentclass[aps,prl,nobalancelastpage,superscriptaddress,twocolumn,longbibliography,nobibnotes]{revtex4-1}
\usepackage{xr}
\usepackage[utf8]{inputenc}
\usepackage[american,british]{babel}
\usepackage[T1]{fontenc}
\usepackage[pdftex]{graphicx}  
\usepackage{xcolor}
\usepackage{dcolumn}
\usepackage{physics}
\usepackage{braket}
\usepackage{bm}
\usepackage{amsmath,amsthm,amssymb}
\usepackage{color}
\usepackage{verbatim}
\usepackage[normalem]{ulem}
\usepackage{dsfont}
\usepackage{hyperref}

\usepackage{subcaption}

\hypersetup{
 colorlinks=true,
 linkcolor=blue,
 anchorcolor = blue,
 citecolor = blue,
 filecolor = blue,
 urlcolor = blue
}

\newcommand{\rrangle}{\rangle\hspace{-0.8mm}\rangle}
\newcommand{\llangle}{\langle\hspace{-0.8mm}\langle}
\newcommand{\kket}[1]
{|{#1}\rrangle }
\newcommand{\bbra}[1]
{\llangle {#1}| }

\def \be {\begin{equation}} 
\def \ee {\end{equation}} 
\def \l {\left(} 
\def \r {\right)} 
\def \la {\langle} 
\def \ra {\rangle}  
\def\lc#1{{ \color{red}  #1}}
\def\bf#1{{ \color{green}  #1}}
\date{}

\newcommand{\saclay}{Universit\'e Paris-Saclay, CNRS, LPTMS, 91405, Orsay, France.}

\begin{document}

\title{
Quantum many-body scars in random unitary circuits
}
\date{\today}

\author{Luca Capizzi} 
\affiliation{\saclay}

\author{Beno\^{\i}t Fert\'e} 
\affiliation{\saclay}
\affiliation{Laboratoire de Physique de l'\'Ecole normale sup\'erieure, ENS, Universit\'e PSL, CNRS, Sorbonne Universit\'e, Universit\'e Paris Cit\'e, F-75005 Paris, France}

\begin{abstract}
Quantum many-body scars are rare exceptions to thermalization: they sustain non-thermal stationary states without the protection of any local conservation law, and are generally expected to be fragile. Here we construct an analytically tractable random unitary circuit hosting a single scar, and derive from first principles the thermalization mechanism governing perturbations thereof — described by a picture of fluctuating interfaces. Surprisingly, despite being thermodynamically irrelevant for local observables, the scar leaves a sharp fingerprint in the entanglement dynamics, driving a transition as a function of perturbation strength that is not probed by any local measurement.
\end{abstract}

\maketitle 

\paragraph{Introduction ---}

The emergence of thermalization in closed many-body quantum systems is one of the central paradigms of modern condensed matter physics. It states that the late-time expectation values of local observables in an isolated system are described by thermal states that account for the conservation of energy as the only global charge \cite{Tasaki_1998,Linden_2009,Tasaki_2016}. While usual thermalization is expected to emerge for generic systems, variation thereof are found in the presence of additional local conservation laws: e.g. in integrable systems, where an infinite number of conserved charges is present, relaxation to equilibrium is expected to be described by Generalized Gibbs Ensembles \cite{Ilievski_2015,Essler_2016,cdy-16,bcdf-16}. 

Notable violations of this \textit{standard framework} have emerged over the last decade, particularly within the context of Rydberg atoms. A key discovery is the existence of specific energy eigenstates that violate the Eigenstate Thermalization Hypothesis (ETH) \cite{Deutsch-91,berry-77,Srednicki-99,D_Alessio_2016} and are known as \textit{quantum many-body scars} \cite{Bernier-17,Turner_2018,Serbyn-21, Moudgalya_2022,Chandran-23,Pizzi_2025}; these states have been considered responsible for non-thermalizing behaviour observed in both experimental setups \cite{Bluvstein-20,Kao-21,Jepsen-22,Su-23} and numerical simulations \cite{Ho-19,Surace-20}. Crucially, these scars are not protected by local conservation laws. Instead, their presence typically originates from exotic algebraic structures—either exact or approximate—such as a spectrum generating algebra \cite{Mark-20, Moudgalya_2022} or non-trivial commutant algebras.

While much of the existing literature focuses on the underlying mechanisms and characterization of quantum many-body scars, several questions regarding large-scale dynamical behaviour in their presence remain open. Recent progress was made in Ref. \cite{mgmgc-24}, where a Lindbladian system featuring a single scar was investigated: it was shown that the evolution following a bipartition protocol — in which the scar is in contact with the infinite temperature state - can be effectively described by a membrane picture, similar to that of entanglement spreading \cite{Nahum-17,nvh-18,Nahum-20}. This description is qualitatively distinct from \textit{hydrodynamics}, typically employed to predict spatio-temporal profiles of local observables in the standard framework. Interestingly, this approach suggests a universal mechanism for scar-mediated large-scale phenomena that lead to an eventual thermalization.

To provide a more rigorous theoretical framework, in this work we introduce an analytically tractable random many-body unitary circuit that serves as a paradigmatic model for local quantum systems with embedded scars. Our construction combines the Shiraishi-Mori approach \cite{sm-17} (also employed in Ref. \cite{Logaric-24} for dual unitary circuits) with the framework of random circuits with symmetries. It also displays analytical tractability, which allows us to derive a membrane picture, underlying a large-scale phenomenon associated with local observables, and identify a novel entanglement transition, induced by the presence of the scar.


\paragraph{The model ---}
We consider a one-dimensional brickwall circuit \cite{Nahum-17}, generated by a 2-sites random unitary gate, with a local Hilbert space of dimension $q$ (spanned by the basis $\{\ket{j}\}_{j=0,\dots,q-1}$). The local gate $u$ is chosen as
\begin{equation}\label{eq:2_sectors}
u = \begin{pmatrix}
1 & 0 \\
0 & \text{Haar}
\end{pmatrix},
\end{equation}
where $u$ acts as the identity on $\ket{00}$ and as a Haar-random unitary on the orthogonal subspace (of dimension $q^2-1$). We consider the associated circuit with $L$ sites ($L$ even for convenience) and open boundary conditions, and we focus on the thermodynamic limit $L\rightarrow \infty$ when possible.

The model in Eq.~\eqref{eq:2_sectors} is motivated by the presence of a stationary state $\ket{00\dots 00}$, which serves as a quantum many-body scar in this context. Meanwhile, the Haar block enables scrambling in the orthogonal sector, driving thermalization toward a unique stationary state. Note that energy is not conserved in unitary circuits, and the infinite-temperature state is the unique stationary state for typical circuits. This contrasts with Hamiltonian dynamics, where the manifold of thermal states is always stationary.

While we consider random circuits for analytical tractability, we expect the phenomenology described here to extend to deterministic circuits hosting a single scar (e.g., those in Res.~\cite{Logaric-24}). Continuous-time evolution should not introduce additional difficulties, except for deterministic Hamiltonian dynamics where energy is conserved. We also emphasize that, although this construction is technically similar to circuits with conserved charges \cite{rpv-18,Khemani-18,Foligno-25}—where diffusive hydrodynamics of local charges were observed—the resulting physics is conceptually distinct: here, the decoupled sector generated by the scar is not protected by any local conservation law.

First, we study the averaged evolution of a local observable $\mathcal{O}$, denoted by $\mathbb{E}[\bra{\psi}\mathcal{O}(t)\ket{\psi}]$. Its description is effectively encoded by a circuit, known as 1-replica model \cite{Nahum-17}, acting on the operators $\mathcal{H}\otimes \mathcal{H}^{*}$, with $\mathcal{H},\mathcal{H}^{*}$ denoting the many-body Hilbert space $(\mathbb{C}^{q})^{\otimes L}$ and its dual, respectively: the associated local generator is the gate
\be\label{eq:1replica_Phi}
\Phi := \mathbb{E}[u\otimes (u^{-1})^{*}],
\ee
with $u^{*}$ being the operator dual to $u$ (represented, in the dual basis, by the matrix transposed to that of $u$). It is not hard to show \cite{rpv-18} that \eqref{eq:1replica_Phi} is a projector, having the identity operator (on two sites) and $\ket{00}\bra{00}$ as eigenvectors with eigenvalue $1$. We represent the local infinite-temperature state acting on a single site (that is, the normalized identity $\mathds{1}/\text{Tr}(\mathds{1}) = \mathds{1}/q$) with the symbol $\bullet$, while $\ket{0}\bra{0}$ is denoted by $\circ$.
The dynamics is projected onto configurations of $\bullet,\circ$ after the first time-step, and the evolution rules induced by \eqref{eq:1replica_Phi} are
\be\label{eq:t_update}
\begin{split}
\kket{\circ \circ} \rightarrow \kket{\circ \circ}, \qquad \kket{\bullet \bullet} \rightarrow \kket{\bullet \bullet},\\
\quad \kket{\bullet \circ},\kket{\circ\bullet } \rightarrow \frac{1}{q+1}\kket{\circ \circ}+\frac{q}{q+1}\kket{\bullet \bullet}.
\end{split}
\ee
This circuit can be viewed as a \textit{quantum channel} (e.g., a Lindbladian evolution in discrete time), that is, a dynamical map of density matrices, represented as vectors using the double-bracket notation. This allows us to draw analogies with Ref.~\cite{mgmgc-24} and study the bipartition protocol between the infinite-temperature state and the scar as the evolution of $\kket{\bullet \dots \bullet \circ \dots \circ}$. Using the rules in Eq.~\eqref{eq:t_update}, we find that the interface between $\bullet$ and $\circ$ behaves as a driven random walker (where the probability of hopping to the right and left is $q/(q+1)$ and $1/(q+1)$, respectively). In the End Matter (EM) we compute the \textit{velocity} and the \textit{diffusion coefficient}, defined as the growth rate of the mean and the variance of the interface, respectively, as
\begin{equation}\label{eq:v_D_anal}
    v = \frac{q-1}{q+1}, \quad \mathcal{D} = \frac{4q}{(1+q)^2}.
\end{equation}
We observe that $v > 0$ and that the infinite-temperature state tends to progressively erase the scar state; this process occurs with the maximum allowed velocity ($v = 1$) when $q \to \infty$. 

To probe the process above, it is sufficient to choose a local observable $\mathcal{O}$ which distinguishes the two stationary states. In particular, using the asymptotic Gaussian behaviour of the random walk, we express the spatiotemporal profile of $\la \mathcal{O}(x,t)\ra$, resulting from the bipartition protocol, as
\be
\label{eq:profile}
\la \mathcal{O}(x,t)\ra \simeq F\l \frac{x-vt}{\sqrt{\mathcal{D} t}} \r \la \mathcal{O}\ra_{\circ} +  \left[1-F\l \frac{x-vt}{\sqrt{\mathcal{D} t}} \r\right] \la \mathcal{O}\ra_{\bullet},
\ee
with $F$ the function
\be
F(z) = \int^{z}_{-\infty}\frac{du}{\sqrt{2\pi}} \ \exp\l -\frac{u^2}{2}\r
\label{eq:erf:rescaling}
\ee
and $\la\mathcal{O}\ra_{\bullet,\circ}$ the expectation value of $\mathcal{O}$ in the infinite temperature state/scar, respectively: this universal profile, equivalent to that found in Ref. \cite{mgmgc-24}, is one of the universal results coming from the membrane picture.

\begin{figure}[t]
\centering \includegraphics[width=0.9\columnwidth]{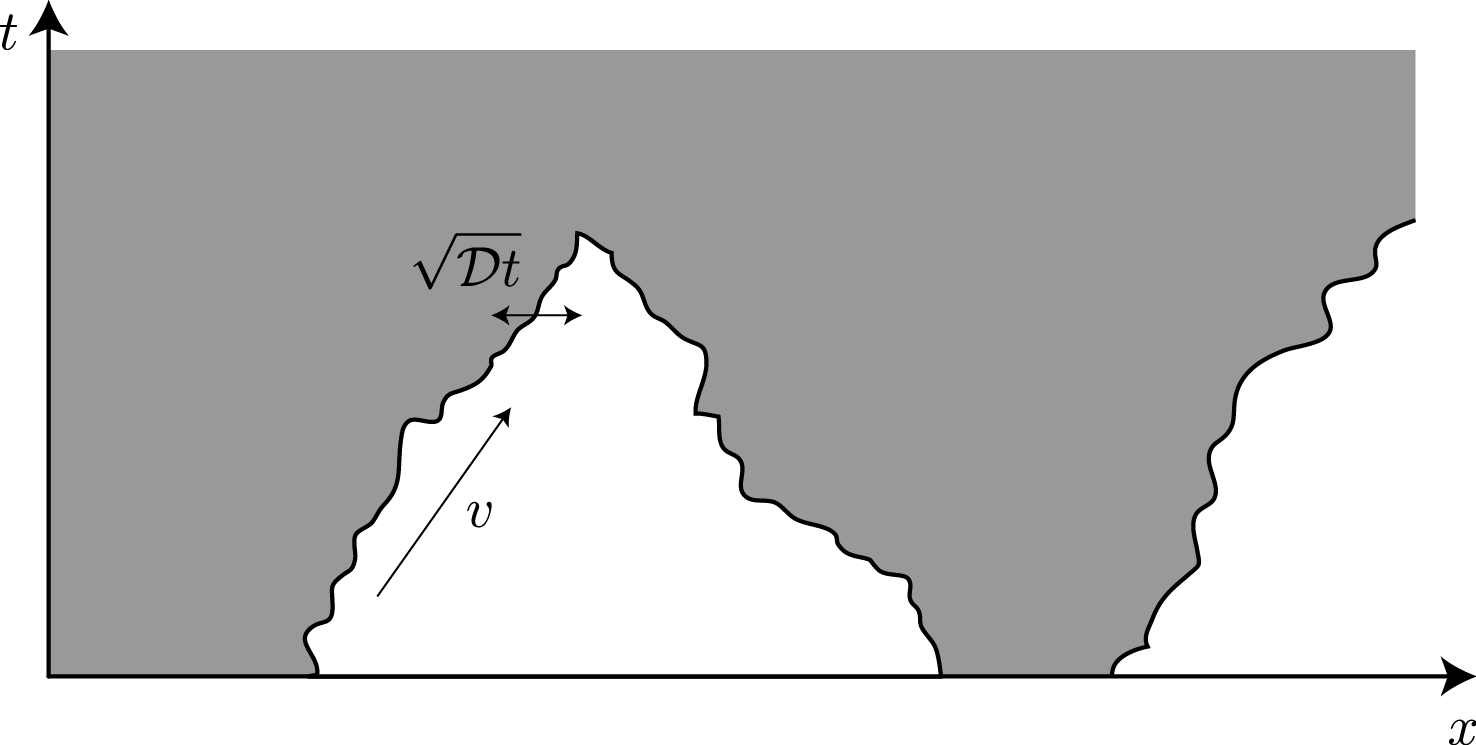}
 \caption{Sketch of the dynamics arising from the stochastic process defined by \eqref{eq:t_update}. Two interfaces propagate stochastically with mean velocities $\pm v$ and diffusion $\mathcal{D}$, undergoing mutual annihilation upon collision.}
 \label{fig:Interface}
\end{figure}

For different initial states, the exact dynamics becomes more difficult to study; however, it can always be described in terms of multiples interfaces that either moves stochastically as random walkers or annihilates with neighboring interfaces, according to the rules in Eq. \eqref{eq:t_update}. Processes with these features, such as the Glauber-Ising model at zero temperature, have been studied in the literature, belong to the so-called \textit{compact directed percolation} universality class, and can be solved exactly with integrability techniques (see Ref. \cite{Henkel-08} for further details). In Fig. \ref{fig:Interface} we provide a sketch which summarizes the salient features of this model. 

\paragraph{Global perturbation---}

In this section we study the dynamics of the initial state
\be\label{eq:psi_state}
\ket{\psi} = (\sqrt{1-\lambda^2}\ket{0}+\lambda \ket{1})^{\otimes L}, \quad \lambda \in [0,1],
\ee
interpreted as a global perturbation of the scar controlled by the perturbation parameter $\lambda$. Our goal is to show that, in the thermodynamic limit $L \to \infty$, the state always converges - at the level of local observables - to the infinite-temperature state as soon as $\lambda \neq 0$. This property corresponds to the instability of the scar under perturbation and is compatible with the thermalization mechanism discussed for the bipartition protocol.

To do so, it is particularly convenient for technical reasons to evolve the operators, rather than the states as in the previous section. We denote by $\bbra{\circ}$ the local operator $\ket{0}\bra{0}$, and by $\bra{1}$ the identity operator $\mathds{1}$ on a single site. In this basis, the evolution rules are
\be\label{eq:t_update_bra}
\begin{split}
\bbra{\circ \circ} \rightarrow \bbra{\circ \circ}, \quad \bbra{11} \rightarrow \bbra{11},\\
\quad \bbra{1 \circ},\bbra{\circ 1 } \rightarrow \frac{q}{q+1}\bbra{\circ \circ}+\frac{1}{q+1}\bbra{11}.
\end{split}
\ee
These rules are similar to those in Eq.~\eqref{eq:t_update}: the crucial difference is that the right/left hopping probability are reversed, and therefore regions with sequences of $\circ$ tend to grow. This counterintuitive discrepancy originates exclusively from the different normalizations of $\bbra{1}$ w.r.t. $\bbra{\bullet}$, which is useful in the context of operator dynamics.

We focus on the evolution of the local observable
\begin{equation}\label{eq:O}
\mathcal{O}(x) := \mathds{1} \otimes \ket{0}\bra{0}_x \otimes \mathds{1},
\end{equation}
which acts as a \textit{local order parameter} to distinguish the two stationary states. In the folded picture, it is represented by $\bbra{\dots 11\circ 11\dots}$ (with $\circ$ at position $x$); in particular, $\la \mathcal{O}\ra_{\circ} = 1$ and $\la \mathcal{O}\ra_{\bullet} = 1/q$.
We evolve the operator according to the rules in Eq.~\eqref{eq:t_update_bra} and compute the overlap with $\ket{\psi}\otimes \ket{\psi}^*$, representing the initial density matrix. By doing so, we obtain
\begin{equation}\label{eq:O_global_pert}
\la \mathcal{O}(t)\ra = \mathbb{E}\left[(1-\lambda^2)^{\# \circ \text{ at time } t}\right].
\end{equation}
The quantity mapped above can be regarded as a large-deviation problem for a pair of random walkers with annihilation; here, we are interested in specific limits.

For instance, proving thermalization for $\lambda \neq 0$, namely
\begin{equation}
\lim_{t \to \infty} \la \mathcal{O}(t)\ra = \la \mathcal{O}\ra_{\bullet},
\end{equation}
is equivalent to showing that the annihilation probability is precisely $1/q$; this is proven in the EM. Furthermore, in the large-$q$ limit, the two walkers move deterministically with opposite velocities ($v=\pm 1$), leading to
\begin{equation}
\la \mathcal{O}(t)\ra = (1-\lambda^2)^{2t+1}, \quad q \to \infty.
\end{equation}
Finally, in the limit of small $\lambda$ (with $\lambda^2t$ held fixed), one can replace the size of the region containing $\circ$ with its average value $2vt$ in Eq.~\eqref{eq:O_global_pert} for those events where the two walkers do not annihilate. This leads to the following asymptotic behavior
\begin{equation}\label{eq:relax_t}
\la \mathcal{O}(t)\ra - \la \mathcal{O}\ra_{\bullet} \asymp \exp\left( -2v\lambda^2t + \dots \right).
\end{equation}
This remarkable result—specifically a relaxation rate of $O(\lambda^2)$ for the globally perturbed scar—is reminiscent of Fermi's Golden Rule and is compatible with previous numerical observations \cite{Lin-20}.

\begin{figure}[t]
\centering
 \includegraphics[width=1\columnwidth]{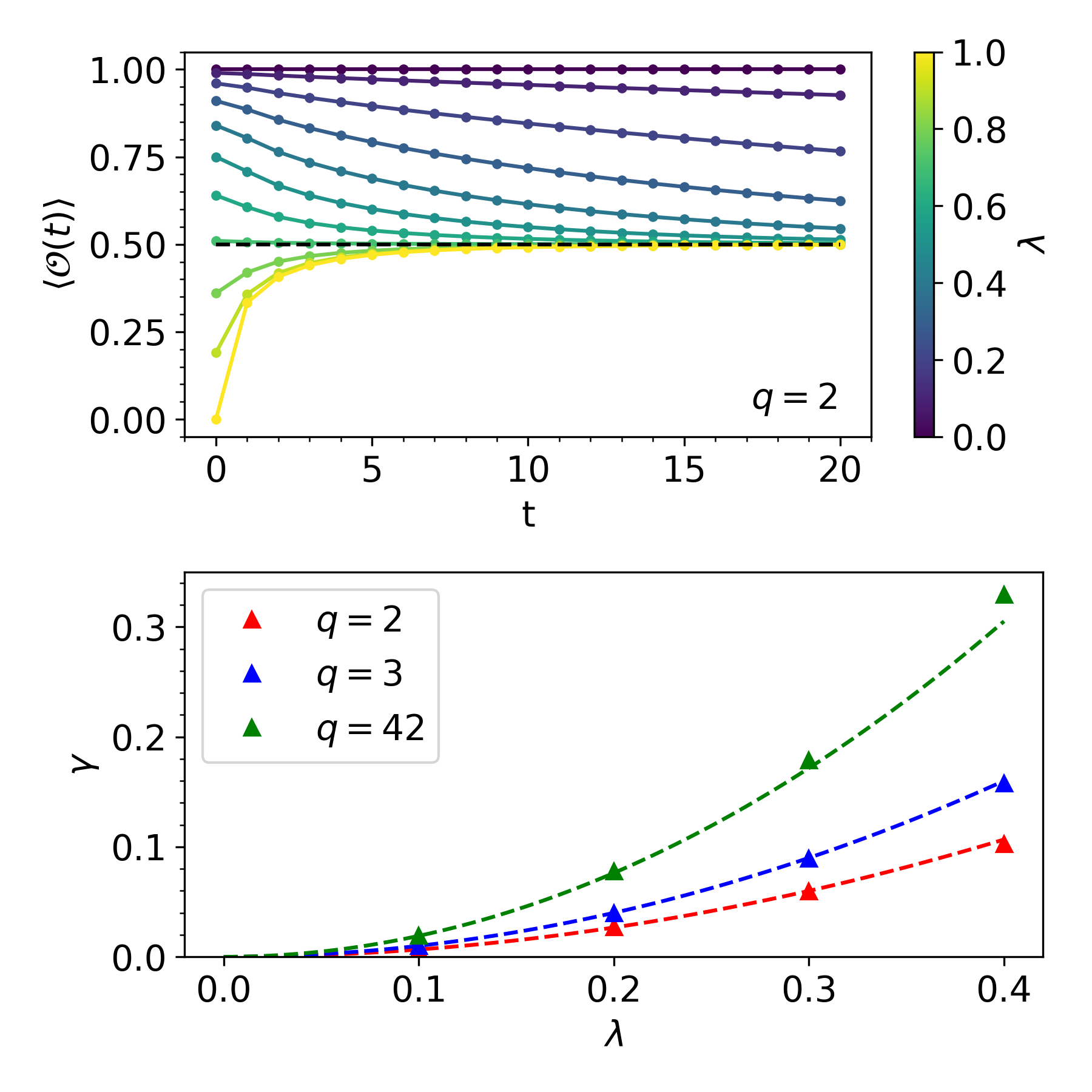}
 \caption{Top panel: Time evolution of the order parameter \eqref{eq:O} for the initial state \eqref{eq:psi_state}, for $q=2$ and some values of $\lambda$. The dashed line represents the infinite-time plateau. Bottom panel: Relaxation rate $\gamma$, as a function of $\lambda$, extracted from an exponential fit. The dashed line represents the analytical prediction at order $O(\lambda^2)$ coming from Eq. \eqref{eq:relax_t}.}
 \label{fig:Ord}
\end{figure}

We test our predictions against exact numerical simulations. For instance, we evolve the operator according to Eq.~\eqref{eq:t_update_bra} and track the coefficients generated in the basis of $\{\circ, 1\}$. In the top panel of Fig.~\ref{fig:Ord}, we plot the evolution of $\langle \mathcal{O}(t)\rangle$ for $q=2$, which suggests a convergence to the expected value $1/q$ as $t \rightarrow \infty$ for $\lambda\neq 0$. In the bottom panel, we plot the relaxation rate $\gamma$ as a function of $\lambda$, extracted from a numerical fit of the form $\langle \mathcal{O}(t)\rangle = 1/q + Ce^{-\gamma t}$ for several values of $q$ ($q=2, 3, 42$). We compare these results with the analytical prediction (dashed line) in Eq.~\eqref{eq:relax_t}, valid at order $O(\lambda^2)$, and obtain good agreement.

\paragraph{Entanglement dynamics---}

In this section, we study entanglement properties which quantify proper quantum features of the system, and require to go beyond the 1-replica model. For instance, for technical reasons, we focus on the average purity and the (annealed averaged) second R\'enyi entropy, namely
\be
S_2 := - \log \mathbb{E}[\text{Tr}(\rho^2_A)].
\ee
with $\rho_A$ the reduced density matrix of the state associated with the region $A$. In this case, the quantity of interest can be computed in terms of a circuit describing the evolution of the $2$-replica model, acting on the Hilbert space $(\mathcal{H}\otimes \mathcal{H}^{*})^{\otimes 2}$: here, the local generator of the circuit is 
\be\label{eq:2replica_Phi}
\Phi := \mathbb{E}[(u\otimes (u^{-1})^{*})^{\otimes 2}].
\ee

\begin{figure}[t]
\centering \includegraphics[width=0.71\columnwidth]{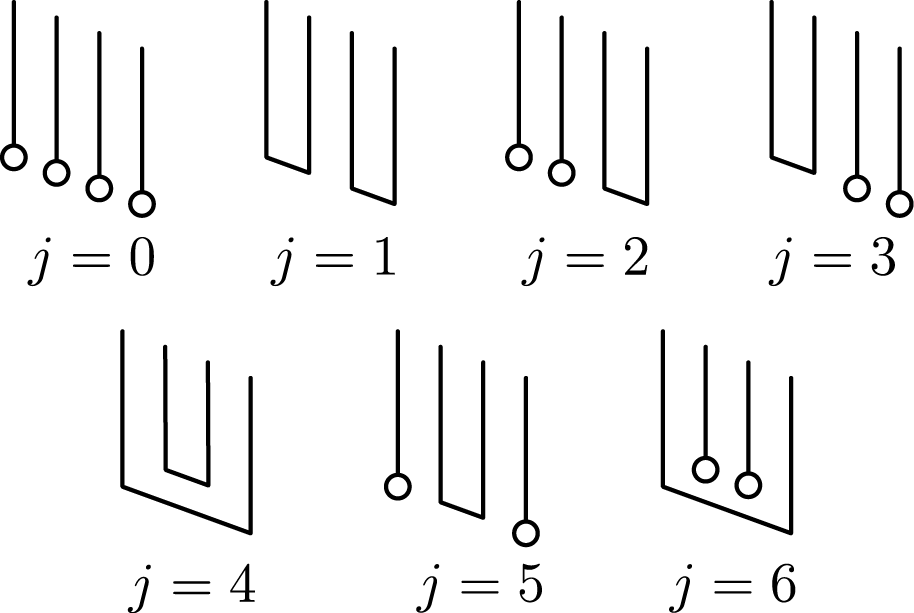}
 \caption{1-site vectors $\{\kket{j}\}_j$ of the 2-replica model. Lines terminating with $\circ$ represents projections on $\ket{0}$, while the other ones represents the identity operators between the corresponding Hilbert spaces.}
 \label{fig:2-replica_basis}
\end{figure}

Similarly to the 1-replica generator \eqref{eq:1replica_Phi}, the 2-replica one is also a projector; its rank is $7$ and it is convenient to find a convenient (non-orthogonal) basis of its image to represent it. These eigenvectors can be obtained by combining the (two) permutation operators, appearing already in the absence of the scar (see e.g. Ref. \cite{fknv-23}), with the projection over the scar. For instance, we consider $\kket{j}_{j=0,\dots,6}$ the one-site vectors represented in Fig. \ref{fig:2-replica_basis} and we compute $\Phi$ as
\be\label{eq:Phi}
\Phi = \sum_{j,j'}M_{jj'}\kket{jj}\bbra{j'j'}.
\ee
Here, $M$ is a $7\times7$ matrix defined by
\be
M = (G \odot G)^{-1}, \quad G_{jj'} := \bbra{j}j'\rangle\hspace{-0.8mm}\rangle,
\ee
with $\odot$ the element-wise product; this definition ensures that $\kket{jj}$ are invariant under $\Phi$.

\begin{figure}[t]
\centering
 \includegraphics[width=1\columnwidth]{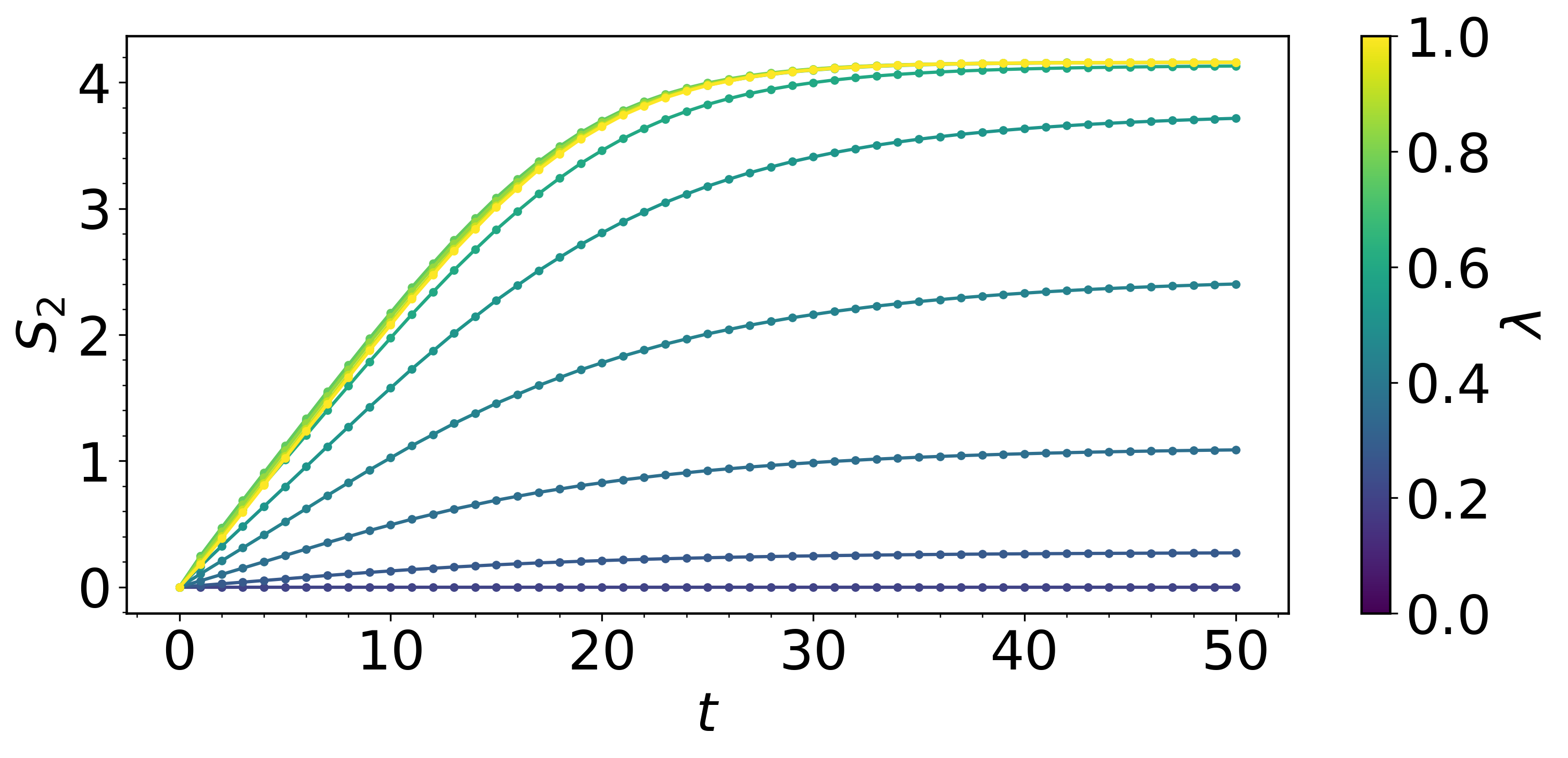}
    \caption{Half-chain R\'enyi entropy $S_2$ of the state \eqref{eq:psi_state} as a function of time, for different values of $\lambda$ ($q=2$). It first grows linearly and then converges toward a $\lambda$-dependent value.}
 \label{fig:S2_as_t}
\end{figure}

In this formalism, the second R\'enyi entropy of a state $\rho$ is computed as:
\be
S_2 = -\log \bbra {\text{SWAP}_A, \mathds{1}_{\bar{A}}} \rho\otimes \rho \rangle\hspace{-0.8mm}\rangle, \quad 
\ee
where $\mathds{1}_{\bar{A}}$ and $\text{SWAP}_A$ refer to the basis elements $j=1,4$ applied to the subspaces $\bar{A}$ and $A$, respectively. We compute numerically the half-chain entropy as a function of time $t$ for the initial state in \eqref{eq:psi_state} and different values of $\lambda$. The results are shown in Fig. \ref{fig:S2_as_t} for $q=2$ and a finite system size $L=14$. For all values of $\lambda$, we observe an initial linear growth in $t$ followed by a crossover toward a plateau. A striking feature is the behavior of both the growth rate and the saturation value as a function of $\lambda$: while they initially increase for small $\lambda$, they reach a plateau for larger values. 

\begin{figure}[t]
    \centering    \includegraphics[width=1\columnwidth]{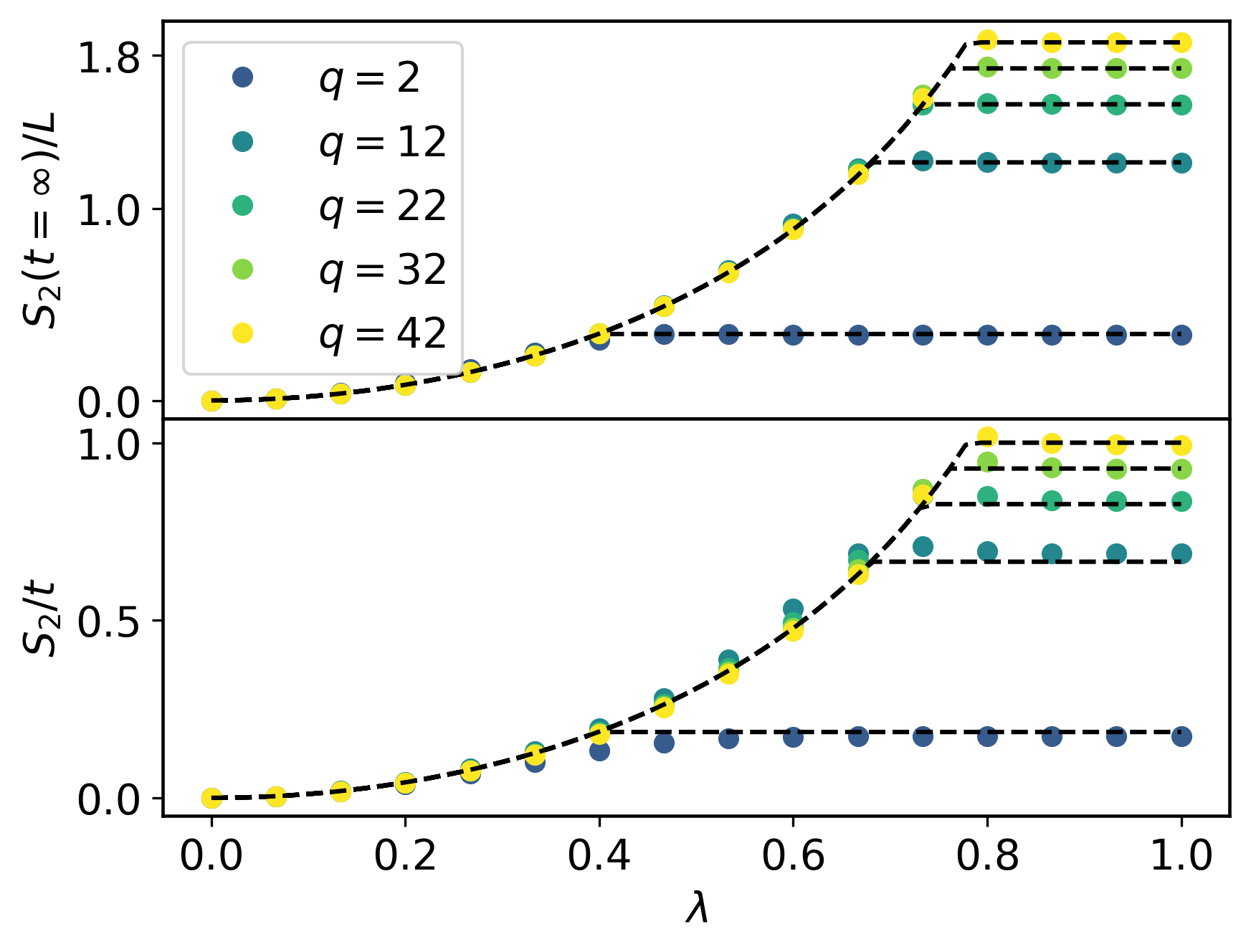}
    \caption{Top: Plateau value of $S_2/L$ for different values of $q$ as a function of $\lambda$. The analytical prediction \eqref{eq:anal_S2_L}, obtained in the thermodynamic limit, is the black dashed line. Bottom: Growth rate of $S_2$. We plot the curve in Eq.~\ref{eq:anal_S2_L} as a dashed line, rescaled by a fitted proportionality constant, finding compatibility.}
\label{fig:combined_S2}
\end{figure}

We claim that the change in behavior above becomes a sharp transition in the thermodynamic limit $L \rightarrow \infty$ for a critical value $\lambda = \lambda^*$ which depends on $q$. We extract the growth from a linear fit $S_2 = at+b$ (performed for $t \in [0,20]$, with $L=12$) for different values of $\lambda$ and $q$; similarly, we extract the density of entropy, denoted by $S_2/L$ for simplicity, from the linear fit $S_2 = a L + b$ with $t=50$ and $L \in [6,8,10,12]$. We plot these two quantities in Fig. \ref{fig:combined_S2}, in the bottom/top panel respectively.

An analytical prediction for the plateau value of $S_2$ can be provided; this is done by replacing the local circuit with a global random matrix having $\ket{0\dots 0}$ as a scar: details are discussed in the EM. The final result, in the limit $L\rightarrow \infty$, is
\be\label{eq:anal_S2_L}
S_2/L = \text{min}\{-2\log(1-\lambda^2), 1/2\log q\},
\ee
implying a transition at $\lambda = \lambda^{*} := \sqrt{1-q^{-1/4}}$. This result is surprising for two reasons. First, it implies a qualitative change in the structure of non-local correlations that are not captured by local observables. Second, in the $q\rightarrow \infty$ limit—associated with the thermalization rate (see Fig. \ref{fig:Ord})—the effect of the scar persists at the level of entanglement for any value of the perturbation $\lambda$, since $\lambda^{*}\rightarrow 1$. Finally, we observe that, although we do not have analytical prediction for the growth rate, a similar behaviour as a function of $\lambda$ is observed numerically: we compare the data with the, properly rescaled, curve in Eq. \eqref{eq:anal_S2_L}, finding good agreement.

\paragraph{Higher order correlators---}

In this section, we study the (averaged) 4-point function 
\be\label{eq:4-point}
\mathbb{E}[\la \mathcal{O}\mathcal{O}(t) \mathcal{O}\mathcal{O}(t)\ra],
\ee
also known as \textit{out-of-time-order correlator} (OTOC) \cite{Roberts-17}, evaluated in the infinite temperature state, for the local observable \eqref{eq:O}. This is an important quantum diagnostic as a probe of \textit{quantum ergodicity} in the sense of Refs. \cite{Aravinda_2021,Fritzsch-25}, formalizing the ideas underlying ETH \cite{Pappalardi-22}. Namely: local observables (at time $t=0$) are \textit{freely-independent} with those evolved at $t \rightarrow \infty$. 

\begin{figure}[t]
    \centering    \includegraphics[width=1\columnwidth]{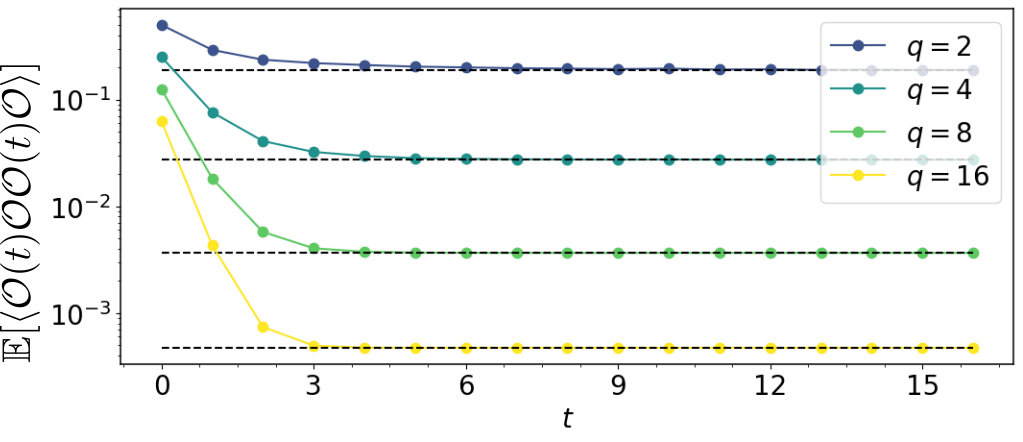}
    \caption{Averaged OTOC $\mathbb{E}[\la \mathcal{O}\mathcal{O}(t) \mathcal{O}\mathcal{O}(t)\ra]$ evaluated in the infinite-temperature state for some values of $q$. The dashed black line is the analytical prediction \eqref{eq:OTOC_pred}.}
\label{fig:OTOC_as_t}
\end{figure}

The value of \eqref{eq:4-point} can expressed in terms of $2$-replica quantities, and its dynamics is traced back to the local circuit generated by $\Phi$ in Eq. \eqref{eq:Phi} (see Ref. \cite{vrps-18}). We compute numerically \eqref{eq:4-point} with exact methods in the thermodynamic limit $L\rightarrow \infty$, and we plot the data, for some values of $q$, in Fig. \ref{fig:OTOC_as_t}. We observe a $q$-dependent plateau, that is consistent with the prediction
\be\label{eq:OTOC_pred}
\underset{t\rightarrow \infty}{\lim} \mathbb{E}[\la \mathcal{O}\mathcal{O}(t) \mathcal{O}\mathcal{O}(t)\ra] = \frac{2q-1}{q^4},
\ee
obtained by assuming quantum ergodicity: details are given in the EM. We also point out that all codes used to produce $2$-replica figures are available at https://github.com/Tyoben/circuit-scars/tree/main.

\paragraph{Conclusions---}
In this work, we investigated a minimal random unitary circuit hosting a single quantum many-body scar, defined as a many-body stationary state not protected by extensive symmetries, following the approach of Ref. \cite{mgmgc-24}. We derive from first principles that the scar is unstable under both global perturbations and bipartition protocols, leading to the eventual emergence of the infinite-temperature state. Furthermore, we provide a large-scale description of these underlying processes. While this suggests the thermodynamic irrelevance of the scar for local observables, we find, counterintuitively, that the entanglement dynamics are highly sensitive to its presence. Specifically, we identify a transition governed by the strength of the global perturbation. We note that conceptually related features appear in the random permutation circuits of Ref. \cite{Schagrin-26}; we propose that this similarity arises because the state $\left(\sum_{j=0}^{q-1} |j\rangle\right)^{\otimes L}$ plays a role analogous to a scar in that context.

The model discussed in this manuscript allows for a straightforward generalization to higher dimensions. We expect the corresponding dynamics of the fluctuating interface to exhibit the features of the Kardar–Parisi–Zhang universality class, similar to conjectures made for higher-dimensional random unitary circuits \cite{Nahum-17}. However, we are not yet aware of techniques to derive this from first principles and hope to return to this problem in future work.

While our focus remained on a minimal scenario with a single scar, similar to that of Ref. \cite{Bocini-24}, it would be interesting to explore settings featuring a tower of scars and dynamical oscillations, as discussed in Ref. \cite{Andrade-24}. Furthermore, we aim to investigate the effect of a small amount of disorder on the PXP automaton (see, e.g., Ref. \cite{Giudici-24}) to determine whether these oscillations persist in the presence of stochasticity. 

\paragraph{Acknowledgements ---}

LC thanks Leonardo Mazza for stimulating discussion on related subjects and motivating the finalization of this work. LC thanks Michele Mazzoni and Gianluca Morettini for their helpful comments on the first version of the manuscript. LC acknowledges support from the ANR project LOQUST ANR-23-CE47-0006-02.

\begin{center}
\begin{large}
\textbf{End Matter}
\end{large}
\end{center}

\paragraph{Velocity and diffusion constant: Analytical prediction}

In this section, we explain how to compute the velocity and the diffusion constant, reported in Eq. \eqref{eq:v_D_anal}. We denote the position of the interface $\hat{x}_t$; using the rules in Eq. \eqref{eq:t_update}, associated with the motion of a random walkers, one deduces that the generating function of $\hat{x}_t$ satisfies
\be\label{eq:gen_fun_hatx}
\la \exp\l z(\hat{x}_t-\hat{x}_0)\r\ra = \l \frac{q}{q+1}e^{z}+\frac{1}{q+1}e^{-z}\r^t.
\ee
Assuming w.l.o.g. $\hat{x}_0 = 0$, $v$ and $\mathcal{D}$ are, by definition, identified by
\be\label{eq:gen_fun_hatx1}
\la \exp\l z \hat{x}_t\r\ra \asymp \exp \left[ t\l vz+ \frac{1}{2}\mathcal{D}z^2 + o(z^2)\r\right].
\ee
From the Eqs. \eqref{eq:gen_fun_hatx} and \eqref{eq:gen_fun_hatx1}, one easily obtains \eqref{eq:v_D_anal}.

\paragraph{Absorbing probability}

In this section we discuss the evolution of the distance between the two interfaces, and we compute the probability of the annihilation thereof, dubbed as \textit{absorbing probability}. We denote the state,where the distance among the interfaces is $x$, as $|x)$; then, from Eq. \eqref{eq:t_update_bra}, it is easy to derive the corresponding evolution rules
\be
\begin{split}
|0)\rightarrow &|0), \ |1) \rightarrow \frac{q}{q+1}|2) + \frac{1}{q+1}|0)\\
|x) \rightarrow &\frac{q^2}{(q+1)^2}|x+2) + \frac{2q}{(q+1)^2}|x) + \\
&\frac{1}{(q+1)^2}|x-2).
\end{split}
\ee
We call $T$ the operator representing the evolution; it allows to represent the absorbing probability starting from $x=1$ as
\be
\underset{t\rightarrow \infty}{\lim}(0|T^t|1) = \frac{1}{q+1}+ \frac{q}{q+1}\underset{t\rightarrow \infty}{\lim}(0|T^t|2).
\ee
The limit can be computed by projecting to the (normalizable) left/right eigenstates with eigenvalues $1$: those are
\be
|R) := |0), \quad (L| := \sum_{x \text{ even}} \frac{1}{q^{x}}(x|, 
\ee
giving $T^t \simeq |R)(L|$ in the large $t$ limit. After putting everything together, we obtain
\be
\underset{t\rightarrow \infty}{\lim}(0|T^t|1) = \frac{1}{q}.
\ee

\paragraph{2nd R\'enyi entropy: saturation value}

We want to compute the saturation value of the second R\'enyi entropy: a trick to do that is the replacement of the local brickwall circuit, that is difficult to analyse analytically, with a huge random matrix having the same formal structure and representing the same late time properties.

Therefore, we consider a $(dd')\times (dd')$ random matrix with the same structure as \eqref{eq:2_sectors}, acting on the Hilbert space $\mathbb{C}^{d}\otimes \mathbb{C}^{d'}$. This decomposition is associated with the bipartition $A \cup \bar{A}$, where thedimensions are $d=q^\ell$ and $d' = q^{L-\ell}$, and $\ell, L-\ell$ are the spatial sizes of $A$ and $\bar{A}$, respectively. In this context, the computation of the 2nd R\'enyi entropy becomes a two-replica calculation with $7\times 7$ matrices, similar to Eq.\eqref{eq:Phi}.

\begin{figure}[t]
\centering \includegraphics[width=0.91\columnwidth]{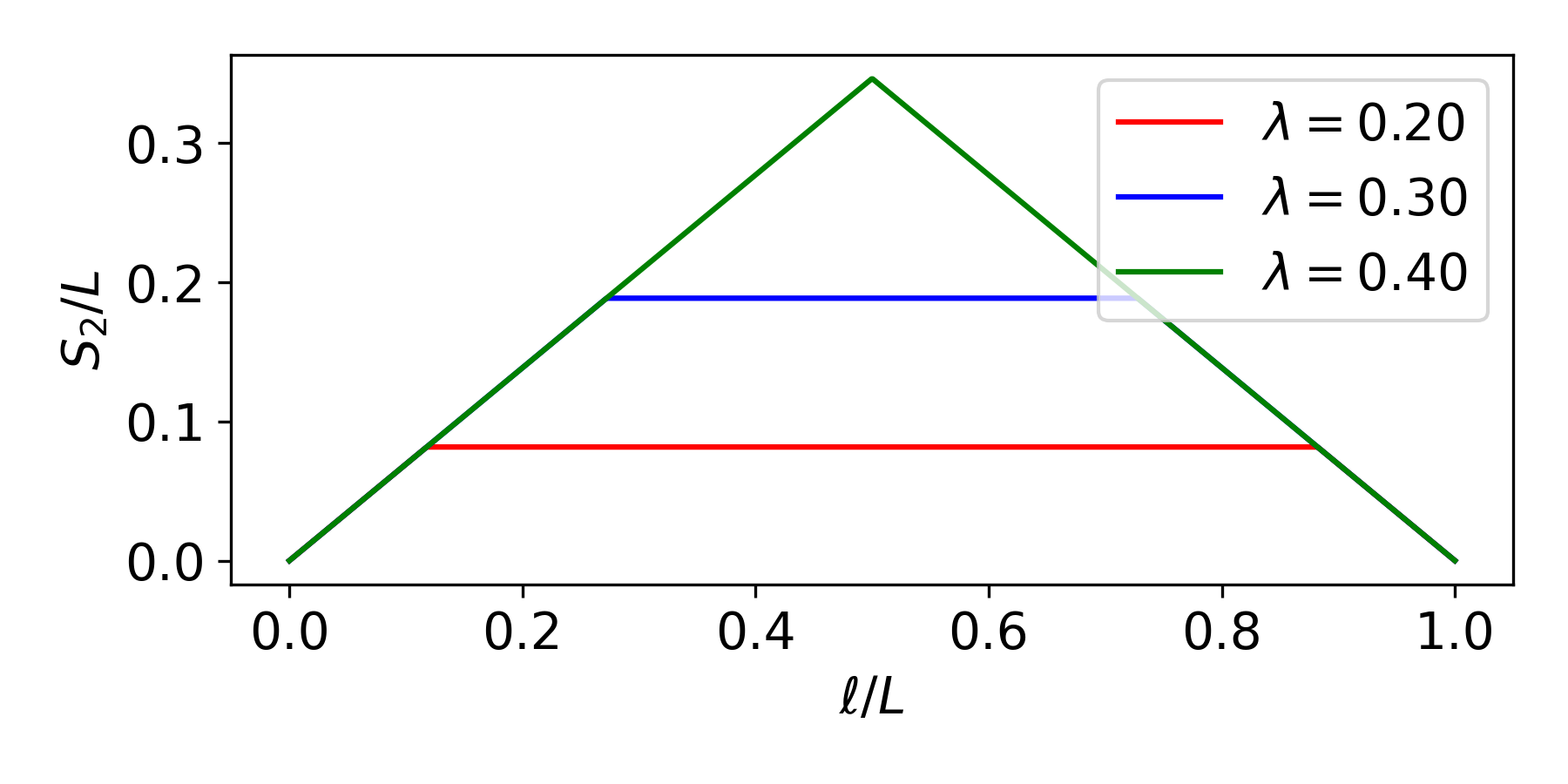}
 \caption{Page curve for the 2nd R\'enyi entropy as a function of $\ell/L$ for $q=2$ and some values of $\lambda$. This is the analytical prediction \eqref{eq:Page}, valid in the thermodynamic limit.}
 \label{fig:Page}
\end{figure}

We define
\be
G(q) := \begin{pmatrix}
1 & 1 & 1 & 1 & 1 & 1 & 1 \\
1 & q^{2} & q & q & q & 1 & 1 \\
1 & q & q & 1 & 1 & 1 & 1 \\
1 & q & 1 & q & 1 & 1 & 1 \\
1 & q & 1 & 1 & q^{2} & q & q \\
1 & 1 & 1 & 1 & q & q & 1 \\
1 & 1 & 1 & 1 & q & 1 & q
\end{pmatrix},
\ee
being the \textit{Gram matrix} $G_{jj'}(q) := \bbra{j}j'\rangle\hspace{-0.8mm}\rangle$, with $\kket{j}$ in Fig. \ref{fig:2-replica_basis}; we consider also
\be
v(q) := \begin{pmatrix}q^{-2\Gamma} \\ 1\\ q^{-\Gamma}\\ q^{-\Gamma}\\ 1 \\ q^{-\Gamma}\\ q^{-\Gamma}
\end{pmatrix},
\ee
with $q^{-\Gamma} := (1-\lambda^2)$ and $\rho$ density matrix of \eqref{eq:psi_state} for $L=1$: this is defined through the overlap $v_j(q) = \bbra{j} \rho\otimes \rho \rangle\hspace{-0.8mm}\rangle$. Then, we introduce $M(d,d') := G(d)\odot G(d')$. After tedious but straightforward algebraic manipulations, we express the averaged purity, after a single application of the huge random unitary matrix
\be\label{eq:purity_sat}
\begin{split}
\mathbb{E}[\text{Tr}(\rho_A)^2] = \sum_{jj'} M_{jj'}(d,d')G_{j4}(d)G_{j1}(d')v(d)_{j'} v(d')_{j'}
\simeq\\
(dd')^{-2-2\Gamma}((dd')^2 + (dd')^{2\Gamma+1}(d+d')),
\end{split}
\ee
where the approximation holds for large $d,d'$.
The expression \eqref{eq:purity_sat} contains two terms inside the parenthesis. Depending on the values of $\ell/L,\Gamma$, one is much smaller than the other one: this lies at the the origin of the transition. Specifically, in the strict limit $L\rightarrow \infty$ with $\ell/L$ kept fixed, we find the so-called \textit{Page curve}\cite{Page-93}
\be\label{eq:Page}
S_2/L = \text{min}\left\{-2\log(1-\lambda^2), \frac{\ell}{L}\log q,\l 1-\frac{\ell}{L}\r\log q\right\},
\ee
which is plotted in Fig. \ref{fig:Page} for $q=2$; this yields \eqref{eq:anal_S2_L} for $\ell/L=1/2$ (half-chain entropy). Eq. \eqref{eq:Page} differs from the usual Page curve due to the presence of the first term inside the min, which is strictly related to the overlap between the state under analysis and the scar.

\paragraph{Plateau of the OTOC: predictions from quantum ergodicity}

We recall the standard textbook definition of free independence between the algebras $\mathcal{A},\mathcal{A}'$ for a given state $\la \dots\ra$ \cite{Voiculescu-92}: given $a_j \in \mathcal{A}$ and $a'_j \in \mathcal{A}'$, generic sets of observables with vanishing expectation value ($\la a_j\ra = \la a'_j\ra =0$), one requires
\be\label{eq:def_qe}
\la a_1 a'_1 a_2 a'_2\dots\ra =0.
\ee
The notion of quantum ergodicity (in many-body systems) is obtained by choosing $\mathcal{A}$ as the algebra of local observables and $\mathcal{A}'$ that of asymptotic observables, obtained by evolving $\mathcal{A}$ at $t=\infty$ \footnote{This is a slight abuse of terminology: technically, we perform the limit $t\rightarrow \infty$ of the expectation value, of the algebra. For instance, it is worth stressing that, by locality, $\mathcal{O}(t) \in \mathcal{A}$ for any finite $t$; therefore, strictly speaking, the algebra obtained with evolution at finite time $t$ coincides always with $\mathcal{A}$.}, and $\la \dots\ra$ a stationary state.

Assuming quantum ergodicity, we first compute $\la [\mathcal{O},\mathcal{O}(t)]^2\ra$ with $\mathcal{O} \in \mathcal{A}$: after a few lines of algebra, we express in the limit $t \rightarrow \infty$
\be\label{eq:comm}
\begin{split}
&\la [\mathcal{O},\mathcal{O}(t)]^2\ra = 2\la [(\mathcal{O}-\la \mathcal{O}\ra)(\mathcal{O}(t)-\la \mathcal{O}(t)\ra)]^2\ra-\\
&-2\la(\mathcal{O}-\la \mathcal{O}\ra)^2(\mathcal{O}(t)-\la \mathcal{O}(t)\ra)^2\ra \rightarrow  -2\la(\mathcal{O}-\la \mathcal{O}\ra)^2\ra^2,
\end{split}
\ee
using the definition \eqref{eq:def_qe} on $\mathcal{O},(\mathcal{O}- \la \mathcal{O}\ra)^2 \in \mathcal{A}$ (with their expectation value being subtracted). Similarly, we compute
\be
\begin{split}
\la \mathcal{O}\mathcal{O}(t)\mathcal{O}\mathcal{O}(t)\ra = &\frac{1}{2}\la [\mathcal{O},\mathcal{O}(t)\ra]^2 \ra + \la \mathcal{O}^2\mathcal{O}(t)^2\ra \rightarrow\\
& - \la (\mathcal{O}-\la \mathcal{O}\ra)^2\ra^2 + \la \mathcal{O}^2\ra^2.
\end{split}
\ee
Finally, we apply the general prediction above to $\mathcal{O}$ in Eq.\eqref{eq:O} and the infinite temperature state and, using $\mathcal{O}^2 = \mathcal{O}$ and $\la \mathcal{O}\ra = 1/q$, we obtain \eqref{eq:OTOC_pred}.

It is worth noting that the concept of quantum ergodicity is distinct from its 
classical counterpart, primarily due to the non-commutative nature of quantum 
observables. Because the \eqref{eq:comm} is explicitly non-vanishing, it is 
erroneous to assume a naive late-time \textit{classicalization} of the dynamics or 
to expect observables to become asymptotically commutative.

\bibliography{bibliography}

\end{document}